\documentclass[11pt]{JHEP3}
\title{Noncommutative Dipole Field Theories And Unitarity}

\author{Dah-Wei Chiou and Ori J. Ganor\\ \\
Department of Physics\\
366 LeConte Hall\\
University of California\\
Berkeley, CA 94720\\ \\
and\\ \\
Lawrence Berkeley National Labs\\
Berkeley, CA 94720\\ \\
Emails: \email{dwchiou,origa@socrates.berkeley.edu}
}

\abstract{We extend the argument of Gomis and Mehen for violation
of unitarity in field theories with space-time noncommutativity to
dipole field theories. In dipole field theories with a timelike
dipole vector, we present 1-loop amplitudes that violate the
optical theorem. A quantum mechanical system with nonlocal
potential of finite extent in time also shows violation of
unitarity.}

\preprint{UCB-PTH-03/28\\ LBNL-53931}
\dedicated{To Hsiao-Ching\\ ---\,DWC}

\usepackage{amsmath}
\usepackage{graphicx}
\usepackage{fancybox}

\newcommand\SUSY[1]{{{\cal N} = 4}} 

\begin{document}


\section{Introduction}\label{sec:intro}

In a seminal paper three years ago \cite{Gomis:2000zz} Gomis and
Mehen demonstrated that certain temporally nonlocal field theories
violate unitarity. The theories studied by Gomis and Mehen were
field theories on a noncommutative space-time
\cite{Fairlie:1988qd}-\cite{Seiberg:1999vs} and their results
agreed with predictions of string theory. In particular, for one
such theory -- $U(N)$ Yang-Mills theory with $\SUSY{4}$
supersymmetry on a noncommutative spacetime -- results from string
theory suggested that unitarity can be restored by adding an
infinite tower of massive fields \cite{Seiberg:2000ms,
Gopakumar:2000na, Barbon:2000sg}. The resulting theory is the
well-known ``Noncommutative Open String Theory'' (NCOS). In fact,
this NCOS is dual to spatially noncommutative $\SUSY{4}$ $U(N)$
Yang-Mills theory \cite{Seiberg:2000ms, Gopakumar:2000na}, which
is unitary. The dual relation between temporally noncommutative
$U(1)$ $\SUSY{4}$ Yang-Mills theory and its spatially
noncommutative version can be demonstrated using field theory
arguments alone \cite{Ganor:2000my}.

Building a theory on a noncommutative spacetime is one way to
realize nonlocality, but it also introduces extra complications
such as the UV/IR relation -- the relation between the momentum of
a particle and its effective transverse size
\cite{Minwalla:1999px}. This makes the nonlocality length scale
increase indefinitely at high momentum \cite{Bigatti:1999iz}.
However, within the framework of string theory, it is possible to
construct other nonlocal theories where the nonlocality scale is
bounded. An example is the dipole field theory (DFT) constructed
in \cite{Bergman:2000cw} (see also
\cite{Cheung:1998te,Banks:1999tr}) and further studied in
\cite{Dasgupta:2000ry}-\cite{Dasgupta}. In these theories the
nonlocality length is a fixed spacetime vector (the
``dipole-vetor''), and DFTs do not exhibit a UV/IR relation. Like
the theories built on a noncommutative space, these DFTs break
Lorentz invariance. Unlike noncommutative space, however, a
suitably constructed DFT can preserve $SO(3)$ rotational
invariance. For this to be the case, the dipole vector has to be
timelike and the theory will be nonlocal in time.

General theories with timelike nonlocality have been studied in
several places (see for instance
\cite{Cheng:2001du}-\cite{Morita:2003vt}) and the question of
unitarity was also addressed. Even the simplest example of a
harmonic oscillator with timelike nonlocality has complex energy
levles. The violation of unitarity in theories on a noncommutative
spacetime has also been recently studied in \cite{ORZ}.

In this work, following Gomis and Mehen, we demonstrate that
timelike DFTs violate unitarity at the 1-loop level. In an
upcoming work, we will explore their NCOS completion to a unitary
theory \cite{Upcoming}.

This paper is organized as follows. In Section \ref{sec:prelim},
we briefly outline the mathematical backgrounds of DFTs which are
originally discussed in \cite{Dasgupta:2001zu}. In Section
\ref{sec:NGDT}, based on \cite{Dasgupta:2001zu}, the
noncommutative dipole gauge theory with adjoint matter is
formulated and the Feynman rules are derived. Later, in Section
\ref{sec:Unitarity}, we show that Feynman diagrams of the theories
with timelike dipole vectors do not satisfy the cutting rule and
thus the unitarity is violated. Finally, in Section
\ref{sec:nlqm}, interested in 0+1D quantum mechanical systems, we
study harmonic oscillators with nonlocal interaction in time and
the result gives complex-valued energy levels or complex poles in
the propagator, both of which indicate the violation of unitarity.
We conclude in Section \ref{sec:discussion} with a discussion of
our results and their relation to the limit of string theory.

\section{Mathematical Backgrounds\label{sec:prelim}}

In DFTs, we assign a constant \emph{dipole vector} $L_{i}$ to each
field
$\phi_{i}$ and define the \textquotedblleft\emph{dipole star product}%
\textquotedblright\ as
\begin{equation}
(\phi_{i}\ast\phi_{j})(x)\equiv\phi_{i}(x-\frac{1}{2}L_{j})\phi_{j}(x+\frac
{1}{2}L_{i}).\label{dipole star}%
\end{equation}
The noncommutativity of the star product originates from the
dipole vector associated to each field. Meanwhile, the requirement
that the dipole star product should be associative requires the
dipole vector of $(\phi_{i}\ast
\phi_{j})(x)$ to be $L_{i}+L_{j}$. The associativity is satisfied via%
\begin{align}
(\phi_{i}\ast\phi_{j})\ast\phi_{k}  & =\left(  \phi_{i}(x-\frac{1}{2}%
L_{j})\phi_{j}(x+\frac{1}{2}L_{i})\right)  \ast\phi_{k}\nonumber\\
& =\phi_{i}(x-\frac{L_{j}+L_{k}}{2})\phi_{i}(x+\frac{L_{i}-L_{k}}{2})\phi
_{k}(x+\frac{L_{i}+L_{j}}{2})\nonumber\\
& =\phi_{i}(x-\frac{L_{j}+L_{k}}{2})(\phi_{j}\ast\phi_{k})(x+\frac{1}{2}%
L_{i})\nonumber\\
& =\phi_{i}\ast(\phi_{j}\ast\phi_{k}).\label{association}%
\end{align}

Similarly to the field theory in noncommutative spacetime, the
natural choice for $Tr$ (trace) is the trace on the matrix values
plus the integral over the spacetime. However, to satisfy the
necessary cyclicity condition, the integrand is restricted to have
a total zero dipole vector. More precisely, the integral serves as
the proper $Tr$ over the functions of zero dipole
vector; i.e.%
\begin{equation}
\int\phi_{1}\ast\phi_{2}\ast\cdots\ast\phi_{n}=\int\phi_{n}\ast\phi_{1}%
\ast\cdots\ast\phi_{n-1}%
\end{equation}
with the condition that $\sum_{i=1}^{n}L_{i}=0$, where $L_{i}$ is
the dipole vector for $\phi_{i}$. As a result, together with the
translational invariance, both the sum of the external momenta and
the total dipole vector should vanish at each vertex.

We also define the complex conjugate of a field. Demanding $(\phi^{\dag}%
\ast\phi)$ to be real, i.e.%
\begin{equation}
\phi^{\dag}\ast\phi=(\phi^{\dag}\ast\phi)^{\dag},
\end{equation}
fixes the dipole vector of $\phi^{\dag}$ to be $-L$ when $\phi$
has dipole vector $L$. Therefore, the dipole vector of any real
(hermitian) field (in
particular, the gauge fields) is zero. As the dipole vector of $\phi^\dag$
is fixed, we can show that%
\begin{equation}
(\phi_{i}\ast\phi_{j})^{\dag}=\phi_{j}^{\dag}\ast\phi_{i}^{\dag}.
\end{equation}

Furthermore, it can be shown that the usual derivative satisfies
the Leibniz rule with respect to the star product, i.e.%
\begin{equation}
\partial_{\mu}(\phi_{1}\ast\phi_{2})=(\partial_{\mu}\phi_{1})\ast\phi_{2}%
+\phi_1\ast(\partial_{\mu}\phi_{2}).
\end{equation}
To formulate the field theory action, therefore, $\partial_{\mu}$
is used as the proper derivative to write down the kinetic terms
of the DFTs.

\section{Noncommutative Dipole Gauge Theory with Adjoint Matter\label{sec:NGDT}}

Equipped with the above mathematics, we are ready to formulate
DFTs. In general, the way to obtain the action of DFTs is to
replace all the products in the actions of the ordinary
commutative field theories with the dipole star products as in
(\ref{dipole star}). Meanwhile, all the fields should be
associated with the proper dipole vectors.

\begin{figure}
\begin{picture}(420,100)(0,15)


 \put(30,70){\scalebox{0.8}{\includegraphics{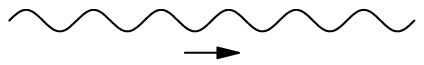}}}
 \put(5,90){(a)}
 \put(20,75){$\mu$}  \put(133,75){$\nu$}  \put(75,60){$k$}
 \put(45,30){\begin{Beqnarray*}
              =\frac{-ig_{\mu\nu}}{k^2+i\varepsilon}
             \end{Beqnarray*}
            }

 \put(165,70){\scalebox{0.8}{\includegraphics{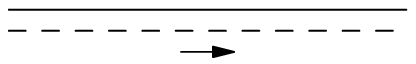}}}
 \put(145,90){(b)}
 \put(210,60){$p$}
 \put(175,30){\begin{Beqnarray*}
               =\frac{i}{p^2-m^2+i\varepsilon}
             \end{Beqnarray*}
            }

 \put(299,40){\scalebox{0.8}{\includegraphics{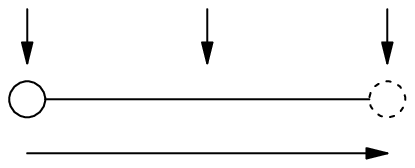}}}
 \put(275,90){(c)}
 \put(295,80){$x-\frac{L}{2}$}  \put(345,80){$x$}  \put(375,80){$x+\frac{L}{2}$}
 \put(285,50){$+g$}  \put(395,50){$-g$}
 \put(345,30){$L$}

\end{picture}
\caption{(a) Propagator of the photon field $A$. (b) Propagator of
the adjoint matter field $\phi$. (c) The dipole structure of
$\phi$ with the positive- and negative-charged ends separated by
$L$.} \label{fig1}
\end{figure}

Here we restrict ourselves to the $U(1)$ gauge theory with the
scalar adjoint matter fields $\phi$ (with the dipole vector $L$).
Other gauge groups can be studied in a similar way. The gauge
fields (photons), being hermitian, should have a zero dipole
vector. Thus, the pure gauge theory is exactly the same as the
commutative theory and the propagator of the gauge field is
unchanged. On the other hand, the gauge fields coupled to the
matter fields reveal the dipole structure of the charged matter
fields. The covariant
derivative of the adjoint matter field is defined as:%
\begin{align}
D_{\mu}\phi  & \equiv\partial_{\mu}\phi+ig(A_{\mu}\ast\phi-\phi\ast A_{\mu
})\nonumber\\
& =\partial_{\mu}\phi+ig\left[  A_{\mu}(x-\frac{L}{2})\phi(x)-\phi(x)A_{\mu
}(x+\frac{L}{2})\right]  ,
\end{align}
and the Lagrangian is
\begin{subequations}
\label{lagrangian}%
\begin{align}
\mathcal{L}  & =D_{\mu}\phi^{\dagger}\ast D^{\mu}\phi
-m^2\phi^\dagger\ast\phi -\frac{1}{4}F_{\mu\nu
}\ast F^{\mu\nu}\\
& =D_{\mu}\phi^{\dagger}\ast D^{\mu}\phi -m^2\phi^\dagger\ast\phi
-\frac{1}{4}F_{\mu\nu}F^{\mu\nu},
\end{align}
which is invariant under the gauge transformation:
\end{subequations}
\begin{subequations}
\begin{align}
\phi & \rightarrow U\ast\phi\ast U^{-1}\\
A_{\mu}  & \rightarrow U\ast A_{\mu}\ast U^{-1}+\frac{i}{g}\partial_{\mu}U\ast
U^{-1}=UA_{\mu}U^{-1}+\frac{i}{g}\partial_{\mu}UU^{-1},
\end{align}
where $U\in U(1)$.

Now, we are ready to study the Feynman rules for the propagators and the vertices.

\subsection{propagators}

In (\ref{lagrangian}), the kinetic term of the gauge field is the
same as the ordinary commutative theory, due to the fact that
$A_{\mu}$ has zero dipole vector. Therefore, the propagator of
$A_{\mu}$ is the same as the commutative QED. [See
Fig.~\ref{fig1}(a).]

Now, we expand the term involving $\phi$:
\end{subequations}
\begin{subequations}
\begin{align}
& D_{\mu}\phi^{\dagger}\ast D^{\mu}\phi\nonumber
=\left(  \partial_{\mu}\phi^{\dagger}-ig\phi^{\dagger}\ast A_{\mu}+igA_{\mu
}\ast\phi^{\dagger}\right)  \ast\left(  \partial^{\mu}\phi+igA^{\mu}\ast
\phi-ig\phi\ast A^{\mu}\right)  \nonumber\\
& =\partial_{\mu}\phi^{\dagger}\ast\partial^{\mu}\phi\label{a}\\
& +ig\left\{  \partial_{\mu}\phi^{\dagger}\ast A^{\mu}\ast\phi-\partial_{\mu
}\phi^{\dagger}\ast\phi\ast A^{\mu}-\phi^{\dagger}\ast A_{\mu}\ast
\partial^{\mu}\phi+A_{\mu}\ast\phi^{\dagger}\ast\partial^{\mu}\phi\right\}
\label{b}\\
& +g^{2}\left\{  \phi^{\dagger}\ast A_{\mu}\ast A^{\mu}\ast\phi-\phi^{\dagger
}\ast A_{\mu}\ast\phi\ast A^{\mu}-A_{\mu}\ast\phi^{\dagger}\ast A^{\mu}%
\ast\phi+A_{\mu}\ast\phi^{\dagger}\ast\phi\ast A^{\mu}\right\}  .\label{c}%
\end{align}
Consider
\end{subequations}
\begin{equation}
(\phi^{\dagger}\ast\phi)(x)=(\phi^{\dagger}\phi)(x-\frac{1}{2}L),
\end{equation}
after integration which becomes
\begin{equation}
\int
d^{D}x\,\partial_{\mu}\phi^{\dagger}\ast\partial^{\mu}\phi=\int
d^{D}x\,\partial_{\mu}\phi^{\dagger}\partial^{\mu}\phi.
\end{equation}
Hence, the dipole dependence in the kinetic term (\ref{a}) as well
as that in the mass term $-m^2\phi^\dagger\ast\phi$ of
(\ref{lagrangian}) are removed. As a result, the propagator of
$\phi$ is exactly the same as the commutative counterpart. [See
Fig.~\ref{fig1}(b). The meaning of the doubled line will be
explained soon.]

\subsection{3-Leg Vertices}

Next, we study the interactions of matter and gauge fields in (\ref{b}), which
correspond to the 3-leg vertices. By the second line of (\ref{association})
and a spacetime translation, the first term in (\ref{b}) becomes%
\begin{align}
(\partial_{\mu}\phi^{\dagger}\ast A^{\mu}\ast\phi)(x) & =\partial_{\mu}%
\phi^{\dagger}(x-\frac{L}{2})A^{\mu}(x-L)\phi(x-\frac{L}{2})\nonumber\\
\text{under translation}  & \rightarrow\partial_{\mu}\phi^{\dagger}(x)A^{\mu
}(x-\frac{L}{2})\phi(x),
\end{align}
and the second term becomes%
\begin{align}
(-\partial_{\mu}\phi^{\dagger}\ast\phi\ast A^{\mu})(x)  & =-\partial_{\mu}%
\phi^{\dagger}(x-\frac{L}{2})\phi(x-\frac{L}{2})A^{\mu}(x)\nonumber\\
\text{under translation}  & \rightarrow-\partial_{\mu}\phi^{\dagger}%
(x)\phi(x)A^{\mu}(x+\frac{L}{2}).
\end{align}
Similar results are obtained for $-\phi^{\dagger}\ast
A_{\mu}\ast\partial^{\mu}\phi$ and $A_{\mu
}\ast\phi^{\dagger}\ast\partial^{\mu}\phi$, and all the terms in
(\ref{b}) are thus equivalent to
\begin{equation}
-igA^{\mu}(x-\frac{L}{2})(\phi^{\dagger}\overleftrightarrow{\partial_{\mu}%
}\phi)(x)+igA^{\mu}(x+\frac{L}{2})(\phi^{\dagger}\overleftrightarrow
{\partial_{\mu}}\phi)(x).
\end{equation}

Each individual term in the above is identical to the 3-leg
interaction in scalar electrodynamics except that the point where
$A^{\mu}$ is coupled is shifted from $x$ to $x\pm L/2$ with the
charge $\mp g$ respectively. This has a natural interpretation in
terms of the dipole structure: when $\phi(x)$ is centered at $x$,
it has two ends with the opposite charges $\pm g$ at $x\mp L/2$.
In the diagram, we use a doubled line to denote the propagator
with the solid and dashed lines representing $+g$ and $-g$
endpoints. [See Fig.~\ref{fig1}(c).]

\begin{figure}
\begin{picture}(420,190)(0,20)


 \put(5,60){\scalebox{0.75}{\includegraphics{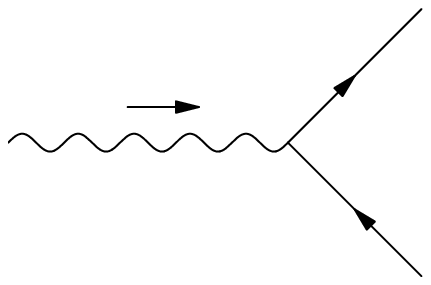}}}
 \put(35,110){$k$}  \put(90,85){$p$}  \put(90,105){$p'$}
 \put(110,95){$\Rightarrow$}

 \put(130,20){\scalebox{0.8}{\includegraphics{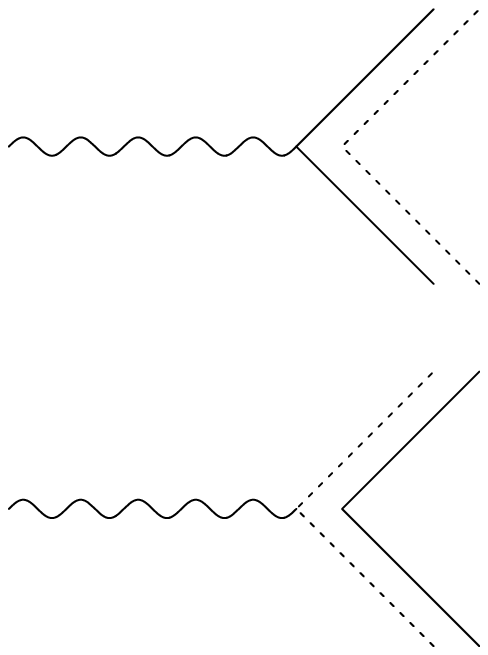}}}

 \put(260,130){\begin{Beqnarray*}
                &=&-ig(p_{\mu}+p'_{\mu}) e^{-\frac{i}{2}k\cdot L} \\
                & &\cdot(2\pi)^D\delta(p-p'+k)
               \end{Beqnarray*}
            }

 \put(260,40){\begin{Beqnarray*}
                &=&+ig(p_{\mu}+p'_{\mu})e^{+\frac{i}{2}k\cdot L}
                \\
                & &\cdot(2\pi)^D\delta(p-p'+k)
               \end{Beqnarray*}
            }

\end{picture}
\caption{Feynman rules for 3-leg vertices.} \label{fig2}
\end{figure}

At the same time, $A^{\mu}$ can couple to $\phi$ at either the
positive-charged or the negative-charged side. The shift of $\pm
L/2$ will introduce an extra phase $e^{\pm ik\cdot L/2}$ in the
momentum space, and the Feynman rules for the vertices are
modified with this \emph{dipole phase}. (See Fig.~\ref{fig2} for
the Feynman rules.)

\subsection{4-Leg Vertices}

Similarly to the second line of (\ref{association}), the star
product of four
fields behaves as%
\begin{align}
(\phi_{i}\ast\phi_{j}\ast\phi_{k}\ast\phi_{l})(x)  & =\phi_{i}(x-\frac
{L_{j}+L_{k}+L_{l}}{2})\phi_{i}(x+\frac{L_{i}-L_{k}-L_{l}}{2})\nonumber\\
& \cdot\phi_{k}(x+\frac{L_{i}+L_{j}-L_{l}}{2})\phi_{l}(x+\frac{L_{i}+L_{j}+L_{k}%
}{2}).
\end{align}
Therefore, the four terms inside the curly bracket of (\ref{c})
are
\begin{align}
& \phi^{\dagger}(x-\frac{L}{2})A_{\mu}(x-L)A^{\mu}(x-L)\phi(x-\frac{L}%
{2})-\phi^{\dagger}(x-\frac{L}{2})A_{\mu}(x-L)\phi(x-\frac{L}{2})A^{\mu
}(x)\nonumber\\
& -A_{\mu}(x)\phi^{\dagger}(x-\frac{L}{2})A^{\mu}(x-L)\phi(x-\frac{L}%
{2})+A_{\mu}(x)\phi^{\dagger}(x-\frac{L}{2})\phi(x-\frac{L}{2})A^{\mu}(x)
\end{align}
and become
\begin{align}
& \phi^{\dagger}(x)A_{\mu}(x-\frac{L}{2})A^{\mu}(x-\frac{L}{2})\phi
(x)-\phi^{\dagger}(x)A_{\mu}(x-\frac{L}{2})\phi(x)A^{\mu}(x+\frac{L}%
{2})\nonumber\\
& -A_{\mu}(x+\frac{L}{2})\phi^{\dagger}(x)A^{\mu}(x-\frac{L}{2})\phi
(x)+A_{\mu}(x+\frac{L}{2})\phi^{\dagger}(x)\phi(x)A^{\mu}(x+\frac{L}{2})
\end{align}
under the spacetime translation.

\begin{figure}
\begin{picture}(420,480)(0,30)


 \put(5,200){\scalebox{0.65}{\includegraphics{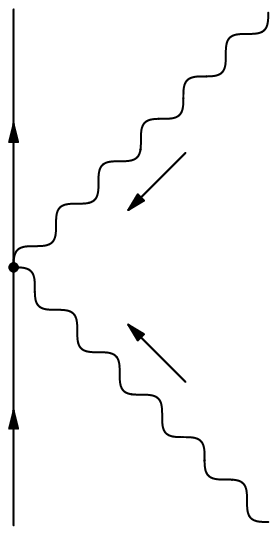}}}
 \put(5,220){$p$}  \put(5,275){$p'$}  \put(75,200){$\mu$}  \put(75,300){$\nu$}
 \put(60,230){$k_1$}  \put(60,280){$k_2$}
 \put(90,250){$\Rightarrow$}

 \put(130,20){\scalebox{0.75}{\includegraphics{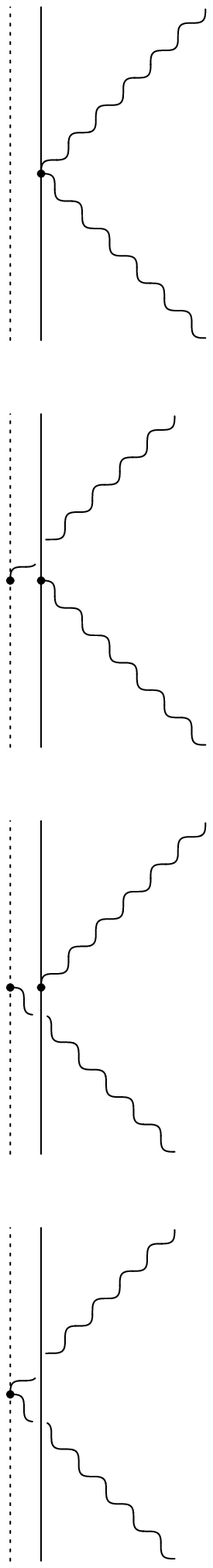}}}

 \put(210,400){\begin{Beqnarray*}
                &=&2ig^2g_{\mu\nu}
                   e^{-\frac{i}{2}(k_1+k_2)\cdot L}\\
                & &\cdot(2\pi)^D\delta(p-p'+k_1+k_2)
               \end{Beqnarray*}
            }

 \put(210,285){\begin{Beqnarray*}
                &=&-2ig^2g_{\mu\nu}
                   e^{-\frac{i}{2}(k_1-k_2)\cdot L}\\
                & &\cdot(2\pi)^D\delta(p-p'+k_1+k_2)
               \end{Beqnarray*}
            }

 \put(210,175){\begin{Beqnarray*}
                &=&-2ig^2g_{\mu\nu}
                   e^{\frac{i}{2}(k_1-k_2)\cdot L}\\
                & &\cdot(2\pi)^D\delta(p-p'+k_1+k_2)
               \end{Beqnarray*}
            }

 \put(210,60){\begin{Beqnarray*}
                &=&2ig^2g_{\mu\nu}
                   e^{\frac{i}{2}(k_1+k_2)\cdot L}\\
                & &\cdot(2\pi)^D\delta(p-p'+k_1+k_2)
               \end{Beqnarray*}
            }

\end{picture}
\caption{Feynman rules for 4-leg vertices.} \label{fig3}
\end{figure}

Again, the two legs of the gauge field can act on either end of
the dipole field $\phi$ at $x\pm L/2$. The Feynman rules for the
4-leg vertices are almost the same as that in scalar
electrodynamics, while the coupling $g^{2}$ is replaced by $(\pm
g)(\pm g)$ and the dipole phases are also introduced. (See
Fig.~\ref{fig3} for the Feynman rules.)

\section{Unitarity of Noncommutative Dipole Field Theory\label{sec:Unitarity}}

\subsection{Optical Theorem and Unitarity}\label{subsec:optical theorem}

In this section, we examine one-loop diagrams of noncommutative
dipole gauge theory to see if unitarity is satisfied.

One straightforward consequence of unitarity is the \emph{optical
theorem}, which states that the imaginary part of any scattering
amplitude arises from a
sum of contributions from all possible intermediate-state particles; i.e.%
\begin{equation}
2\operatorname{Im}M(a\rightarrow b)=\sum_{f}M^{\ast}(b\rightarrow
f)M(a\rightarrow f),\label{optical thm}%
\end{equation}
where the sum runs over all possible sets $f$ of final-state particles and
includes phase space integrations for each particle in $f$.

Quantum field theories actually satisfy more restrictive relation
called the \emph{cutting rule}, which is a generalization of the
optical theorem to Feynman diagrams of all orders in perturbation
theory. This states that the imaginary part of any Feynman diagram
is given by the following algorithm:

\begin{enumerate}
\item Cut through the diagram into two separate pieces in all possible ways
such that the cut propagators can simultaneously be put on shell.

\item For each cut, place the virtual particle on-shell by replacing the
propagator with a delta function:%
\begin{equation}
\frac{1}{p^{2}-m^{2}+i\varepsilon}\rightarrow-2\pi i\delta(p^{2}-m^{2}).
\end{equation}

\item Perform the integrals over $p$ and sum over the contributions of all
possible cuts.
\end{enumerate}

The cutting rule is more general than the constraint of unitarity
because it applies to the scattering amplitudes as well as the
off-shell Green's functions. The detailed discussion can be found
in many QFT textbooks \cite{Peskin:Schroeder}.

As shown in \cite{Gomis:2000zz}, the spacetime noncommutative
scalar field theory does not obey unitarity when there is a
space-time noncommutativity ($\theta^{0i}\neq0$), while in the
case of spatial noncommutativity ($\theta^{0i}=0$,
$\theta^{ij}\neq0$) the cutting rules are satisfied. Following
\cite{Gomis:2000zz}, in our case of the DFT, we will first show
that the two-point function of $\phi$ (self energy diagram)
violates the usual cutting rules when the associated dipole vector
$L$ is \emph{timelike}. On the other hand, when $L$ is
\emph{spacelike}, the unitarity is satisfied. Later, we consider
the $2\phi\rightarrow2\phi$ scattering, and the same conclusion is
made.

However, the diagrams with external photon legs signal no
unitarity violation (at least at one-loop level) even for timelike
$L$. This suggests that there might be some symmetry which keeps
the unitarity in the limit of the string theory, but we will not
explore this further.

The reason for the failure of unitarity is very similar to that of
\cite{Gomis:2000zz}. Since the Feynman rules for the vertices are
manifestly real functions of momenta (typically of the form $\sim
e^{ik\cdot L}+e^{-ik\cdot L}$ ), at first look, one would expect
that the Feynman diagrams could develop a branch cut only when the
internal lines go on-shell. This would imply that the imaginary
parts of Feynman diagrams would be given by the same cutting rules
as the ordinary commutative field theory and thus unitarity would
be satisfied. However, a closer examination of the high energy
behavior of the oscillatory factors that arise from the dipole
phases shows that unitarity is broken. A Feynman integral can be
defined via analytic continuation by Wick rotation, but the
resulting amplitude will develop branch cuts when $L^{2}>0$
(timelike) back in Minkowski space. These additional branch cuts
are responsible for the breakdown of the cutting rule.

For $L^{2}=0$ (lightlike), the amplitude is ill-defined because of
the infrared divergence. A sensible theory should have
infrared-safe observables and this may requires all order
re-summation of infrared divergent terms in the perturbative
series. We will not attempt to address this issue in this paper
and focus on the amplitudes which do not suffer from infrared
singularities. (See \cite{Gomis:2000zz}.)

\subsection{Two-Point Function of $\phi$}

\begin{figure}
\begin{picture}(420,150)(0,0)


 \put(0,50){\scalebox{0.75}{\includegraphics{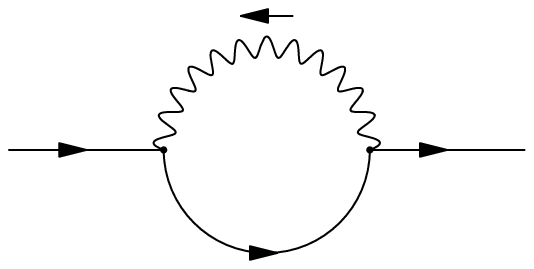}}}
 \put(10,60){$p$}  \put(90,60){$p$}  \put(50,40){$p+k$}  \put(55,110){$k$}
 \put(40,70){$\mu$}  \put(73,70){$\nu$}
 \put(120,70){$\Rightarrow$}

 \put(130,15){\scalebox{0.8}{\includegraphics{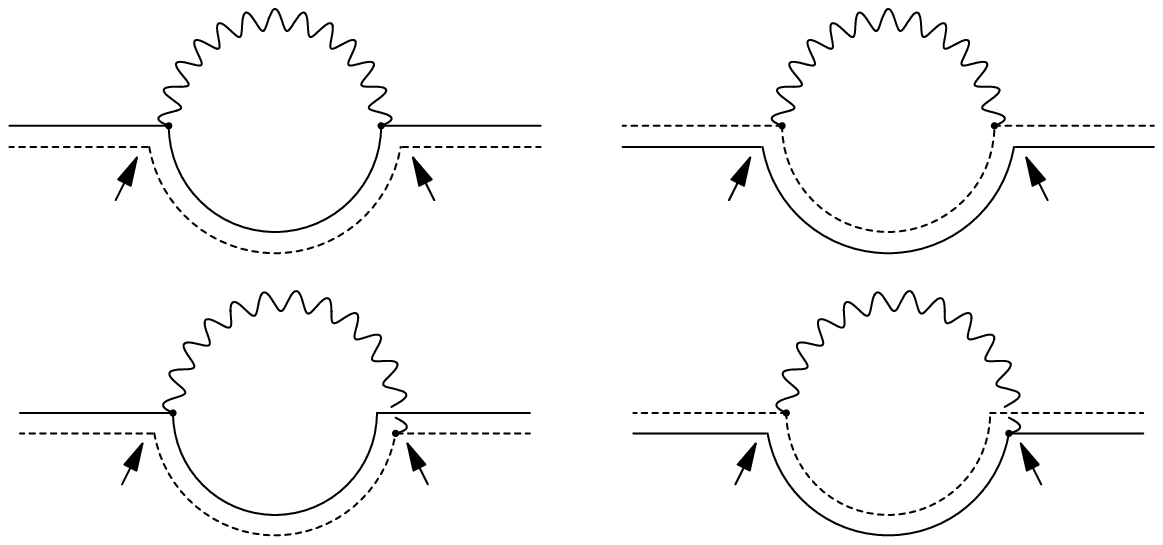}}}

 \put(270,110){+}  \put(270,45){+}

 \put(120,83){$-ige^{-ik\cdot L/2}$}  \put(225,83){$-ige^{ik\cdot L/2}$}
 \put(120,20){$-ige^{-ik\cdot L/2}$}  \put(225,20){$ige^{-ik\cdot L/2}$}
 \put(281,83){$ige^{ik\cdot L/2}$}  \put(370,83){$ige^{-ik\cdot L/2}$}
 \put(281,20){$ige^{ik\cdot L/2}$}  \put(370,20){$-ige^{ik\cdot L/2}$}

\end{picture}
\caption{Feynman diagrams for the two-point function of $\phi$.
The two diagrams on the top are the planar part while those on the
bottom are the nonplanar part. Each vertex contribution with the
corresponding dipole phase is indicated.} \label{fig4}
\end{figure}

Now, we study the two-point functions of $\phi$. The one-loop
Feynman diagrams are listed in Fig.~\ref{fig4}. Since $\phi$ has
the dipole structure and $A $ can act on either end of the dipole,
the diagrams have planar and nonplanar parts.
By the Feynman rules, the planar amplitude is:%
\begin{equation}
iM_{p}=-2g^{2}\int\frac{d^{D}k}{(2\pi)^{D}}\frac{g^{\mu\nu}}{k^{2}%
+i\varepsilon}\frac{(2p_{\mu}+k_{\mu})(2p_{\nu}+k_{\nu})}{(p+k)^{2}%
-m^{2}+i\varepsilon},
\end{equation}
and the nonplanar amplitude is:%
\begin{equation}
iM_{np}  =2g^{2}\int\frac{d^{D}k}{(2\pi)^{D}}\frac{g^{\mu\nu}}%
{k^{2}+i\varepsilon}\frac{(2p_{\mu}+k_{\mu})(2p_{\nu}+k_{\nu})}{(p+k)^{2}%
-m^{2}+i\varepsilon}\cos k\cdot L,
\end{equation}
where $D$ is the spacetime dimension. We will focus on the
nonplanar terms because it is obvious that the planar parts
satisfy the unitarity just as the ordinary commutative field
theory does.

First, we perform a Wick rotation to Euclidean space by the usual
analytic continuation $p^{0}\rightarrow ip_{E}^{2}$,
$p^{2}\rightarrow -p_{E}^{2}$, $L=(L^{0},\vec{L})\rightarrow
L_{E}=(-iL^{0},\vec{L})$, $p\cdot
L\rightarrow p_{E}\cdot L_{E}$:%
\begin{equation}
M_{np}=-g^{2}\int\frac{d^{D}k_{E}}{(2\pi)^{D}}\frac{-(k_{E}+2p_{E})^{2}}%
{k_{E}^{2}[(k_{E}+p_{E})^{2}+m^{2}]}(e^{ik_{E}\cdot L_{E}}+c.c.),
\end{equation}
and then combine the denominators by Schwinger parameters%
\begin{equation}
\frac{1}{k_{E}^{2}[(k_{E}+p_{E})^{2}+m^{2}]}=\int_{0}^{1}dx\int_{0}^{\infty
}d\alpha\,\alpha\,e^{-\alpha xk_{E}^{2}}e^{-\alpha(1-x)[(k_{E}+p_{E}%
)^{2}+m^{2}]}%
\end{equation}
to obtain%
\begin{equation}
M_{np}=-g^{2}\int\frac{d^{D}k_{E}}{(2\pi)^{D}}\int_{0}^{1}dx\int_{0}^{\infty
}d\alpha\,\alpha(k_{E}+2p_{E})^{2}e^{-\alpha xk_{E}^{2}-\alpha(1-x)[(k_{E}%
+p_{E})^{2}+m^{2}]+ik_{E}\cdot L_{E}}+c.c.\label{np1}
\end{equation}

Consider the exponent in (\ref{np1}):
\begin{align}
& \alpha xk_{E}^{2}+\alpha(1-x)[(k_{E}+p_{E})^{2}+m^{2}]-ik_{E}\cdot
L_{E}\nonumber\\
& =\left\{  \alpha\lbrack k_{E}+(1-x)p_{E}-\frac{i}{2\alpha}L_{E}%
]^{2}+x(1-x)p_{E}^{2}\right. \\
& \left.  +(1-x)m^{2}+\frac{L_{E}^{2}}{4\alpha^{2}}+\frac{i(1-x)}{\alpha}%
p_{E}\cdot L_{E}\right\} \\
& \equiv\alpha\left\{  k_{E}^{\prime2}+x(1-x)p_{E}^{2}+(1-x)m^{2}+\frac{L_{E}^{2}%
}{4\alpha^{2}}+\frac{i(1-x)}{\alpha}p_{E}\cdot L_{E}\right\}  .
\end{align}
By a change of variable from $k_E$ to $k'_E$, (\ref{np1}) becomes%
\begin{align}
M_{np}  & =-g^{2}\int\frac{d^{D}k_{E}^{\prime}}{(2\pi)^{D}}\int_{0}^{1}%
dx\int_{0}^{\infty}d\alpha\,\alpha\left(  k_{E}^{\prime}+(x+1)p_{E}+\frac
{i}{2\alpha}L_{E}\right)  ^{2}\nonumber\\
& \cdot e^{-\alpha\lbrack k_{E}^{\prime2}+x(1-x)p_{E}^{2}+(1-x)m^{2}%
+\frac{L_{E}^{2}}{4\alpha^{2}}+\frac{i(1-x)}{\alpha}p_{E}\cdot L_{E}%
]}+c.c.,\label{np2}%
\end{align}
where only $k_{E}^{\prime2}$ and
$[(x+1)p_{E}+\frac{i}{2\alpha}L_{E}]^{2}$ in the first line
contribute due to the spherical symmetry in $k_{E}^{\prime}$
space.
After integrating over $dk_{E}^{\prime}$, (\ref{np2}) becomes%
\begin{align}
M_{np}  & =-\frac{g^{2}}{2^{D}\pi^{D/2}}\int_{0}^{1}dx\int_{0}^{\infty}%
d\alpha\,\alpha\left\{  \frac{D/2}{\alpha^{(D+2)/2}}
+\frac{\left[(x+1)p_E+\frac{i}{2\alpha}L_E\right]^2}{\alpha^{D/2}%
}\right\} \nonumber\\
& \cdot e^{-\alpha\lbrack x(1-x)p_{E}^{2}+(1-x)m^{2}+\frac{L_{E}^{2}}%
{4\alpha^{2}}+\frac{i(1-x)}{\alpha}p_{E}\cdot L_{E}]}+c.c.
\end{align}

In particular, we can work out the integration over $d\alpha$ for $D=3$:%
\begin{align}
 M_{np,D=3} &= \frac{g^{2}}{4\pi}\int_{0}^{1}dx\frac{e^{-\sqrt{(1-x)(m^{2}%
+p_{E}^{2}x)L_{E}^{2}}}}{\sqrt{L_{E}^{2}}}
 \left\{ \rule{0mm}{6mm}-\cos[(x-1)p_{E}\cdot L_{E}] \cdot \right. \nonumber \\
& \left(  2+\frac{\sqrt{L_{E}^{2}}}{\sqrt{(1-x)(m^{2}+p_{E}^{2}x)}}%
[(x-1)m^{2}+p_{E}^{2}(2x^{2}+x+1)]\right) \nonumber \\
& \left. \ \;+\sin[(x-1)p_{E}\cdot L_{E}](2(1+x)p_{E}\cdot
L_{E})\rule{0mm}{6mm}\right\}.
\end{align}
This is very complicated, but it is greatly simplified if we consider the
on-shell condition ($p^{2}=m^{2}$):%
\begin{align}
M_{np,D=3}  & =\frac{g^{2}}{4\pi}\int_{0}^{1}dx\frac{e^{-(1-x)m\sqrt{-L^{2}}}%
}{\sqrt{-L^{2}}} \left\{\rule{0mm}{4mm}\right. \frac{2}{1-x}[m(x^{2}+1)\sqrt{-L^{2}%
}+x-1]\cos((x-1)p\cdot L)  \nonumber\\
&  \left. +2p\cdot L (1+x) \sin((x-1)p\cdot L) \rule{0mm}{4mm}\right\},\label{np3}%
\end{align}
where we have analytically continued back to the Minkowski space.

For $L^{2}<0$ (spacelike), obviously (\ref{np3}) gives $\operatorname{Im}%
M_{np}=0 $. Meanwhile, the right hand side of (\ref{optical thm})
is zero because the process that an on-shell massive particle
decays into an on-shell massive particle plus a real photon is
kinematically forbidden. Therefore, unitarity is satisfied.

On the other hand, for $L^{2}>0$ (timelike),
$\sqrt{-L^{2}}=i\left\vert L\right\vert $ and (\ref{np3}) gives
$\operatorname{Im}M\neq0$, while the right hand side of
(\ref{optical thm}) is still zero. Therefore, we find that
unitarity is violated when the dipole vector $L$ is timelike (at
least for the on-shell condition).

\subsection{$2\phi\rightarrow2\phi$ Scattering Amplitude}

Next, we consider the $2\phi\rightarrow2\phi$ scattering
amplitude. The one-loop Feynman diagrams are shown in
Fig.~\ref{fig5}. For each 4-leg vertex, we have an extra dipole
phase depending on which ends the two virtual photons act on.
Consequently, the total contribution due to the vertices with the
dipole phases is
\begin{align}
& 2i\left(  ge^{\frac{i}{2}k\cdot L}-ge^{-\frac{i}{2}k\cdot
L}\right) \left(  ge^{\frac{i}{2}(s-k)\cdot
L}-ge^{-\frac{i}{2}(s-k)\cdot L}\right)
\cdot\nonumber\\
& 2i\left(  ge^{-\frac{i}{2}k\cdot L}-ge^{\frac{i}{2}k\cdot
L}\right) \left(  ge^{-\frac{i}{2}(s-k)\cdot
L}-ge^{+\frac{i}{2}(s-k)\cdot L}\right)
\nonumber\\
& =-16g^{4}\left(  1-\cos k\cdot L-\cos(s-k)\cdot L+\cos(k\cdot
L)\cos (s-k)\cdot L\right),  \label{phase term}
\end{align}
where $s=p_{1}+p_{2}$.

\FIGURE{
\begin{picture}(420,450)(0,0)


 \put(150,370){\scalebox{0.9}{\includegraphics{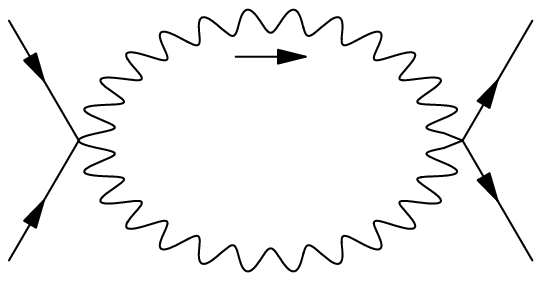}}}
 \put(145,420){$p_1$}  \put(145,390){$p_2$}  \put(220,415){$k$}  \put(70,400){$s=p_1+p_2$}
 \put(220,350){$\Downarrow$}

 \put(0,0){\scalebox{0.9}{\includegraphics{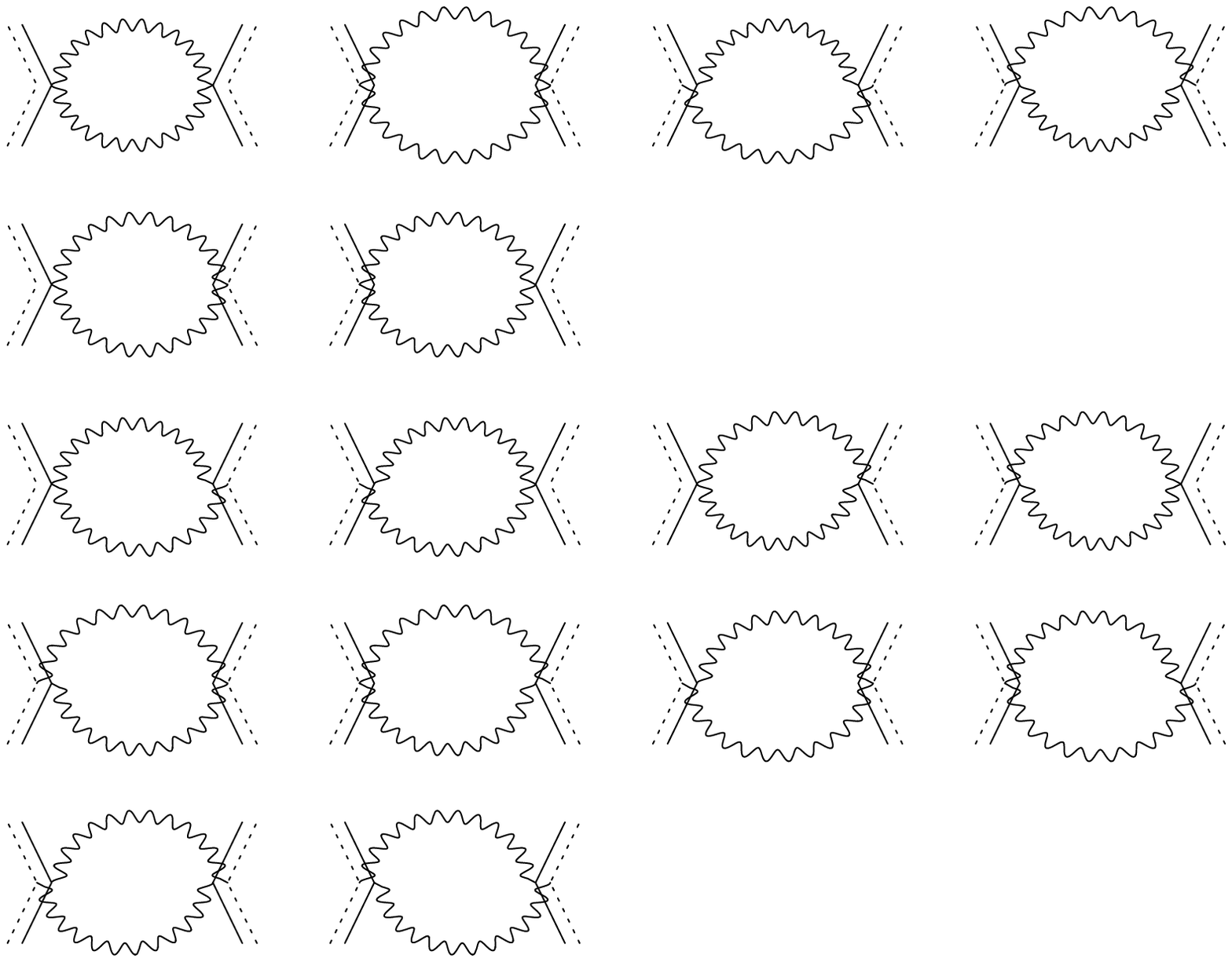}}}
 \put(2,332){(a)}  \put(2,200){(b)}  \put(2,65){(c)}

\end{picture}
\caption{Feynman diagrams for $2\phi\rightarrow 2\phi$ process:
The diagrams of (a) give the vertex contribution
$-4g^4(1+1+1+1+e^{is\cdot L}+e^{-is\cdot L})=-8g^4(2+\cos s\cdot
L)$. Those of (b) give $16g^4(\cos k\cdot L+\cos (s-k)\cdot L)$,
and those of (c) give $-4g^4(e^{-i(s-2k)\cdot L}+e^{i(s-2k)\cdot
L})=-8g^4\cos(s-2k)\cdot L$. All together we get the result in
(\ref{phase term}).} \label{fig5}
}

As shown in Fig.~\ref{fig5}, the intermediate loop is made of
photon propagators, which have no dipole structure, so the
``non-planarity'' is no longer a proper way to describe the
non-triviality. Nevertheless, we can classify the diagrams
according to the way in which the dipole phase is coupled to the
internal momentum $k$ of the loop. In Fig.~\ref{fig5}(a), the
dipole phases are totally decoupled from the internal momentum and
therefore unitarity of those diagrams is guaranteed as the
ordinary commutative theory is unitary. The dipole phases from the
diagrams in Fig.~\ref{fig5}(a) are the ``trivial'' part of
(\ref{phase term}). On the other hand, the diagrams in
Fig.~\ref{fig5}(b)(c) altogether give the nontrivial amplitude\footnote{In fact, the
diagrams in Fig.~\ref{fig5}(a) can all be drawn on the surface of a
sphere although some of them are non-planar, while those in Fig.~\ref{fig5}(b)(c)
can be drawn on the surface of a torus. Therefore, instead of non-planarity,
we can use \emph{genus} to classify the non-triviality in all cases.}
\begin{align}
iM_{nt} =&
\frac{1}{2}g^{4}g_{\mu\nu}g_{\rho\sigma}\int\frac{d^{D}k}{(2\pi)^{D}}
\frac{-ig^{\mu\rho}}{k^{2}+i\varepsilon}\frac{-ig^{\nu\sigma}}{(s-k)^{2}%
+i\varepsilon}\nonumber\\
& \cdot\left\{  16\cos k\cdot L+16\cos(s-k)\cdot
L-8\cos(s-2k)\cdot L\right\},
\end{align}
where the factor $1/2$ is due to the internal bosonic loop.

Following the same procedure in the previous subsection, in
Euclidean space, we have
\begin{align}
M_{nt}
& =\frac{-4D[1+(-1)^{D}]g^{4}}{2^{D-1}\pi^{D/2}}\int_{0}^{1}dx\int_{0}%
^{\infty}d\alpha\,\alpha^{1-D/2}e^{-\alpha\lbrack(x(1-x)s_{E}^{2}+\frac
{L_{E}^{2}}{4\alpha^{2}}]}\cos((1-x)s_{E}\cdot L_{E})\nonumber\\
& +\frac{2Dg^{4}}{2^{D-1}\pi^{D/2}}\int_{0}^{1}dx\int_{0}^{\infty}%
d\alpha\,\alpha^{1-D/2}e^{-\alpha\lbrack(x(1-x)s_{E}^{2}+\frac{L_{E}^{2}%
}{\alpha^{2}}]}\cos((1-2x)s_{E}\cdot L_{E}).\label{np-a}%
\end{align}

\begin{figure}
\begin{picture}(420,80)(0,0)


 \put(25,0){\scalebox{1}{\includegraphics{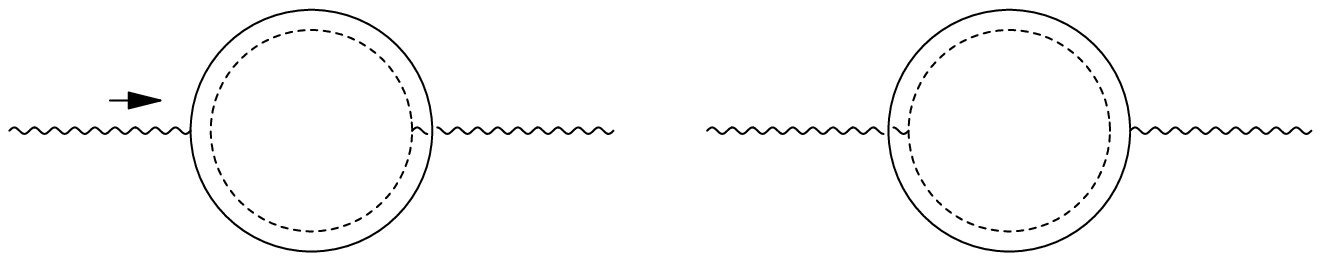}}}
 \put(70,52){$k$}  \put(35,45){$\mu$}  \put(205,45){$\nu$}
 \put(225,35){$+$}

\end{picture}
\caption{Nonplanar one-loop Feynman diagrams for vacuum
polarization.} \label{fig6}
\end{figure}

We evaluate this integral for $D=3$ and $4$. Back to Minkowski
space, (\ref{np-a}) gives
\begin{subequations}
\label{np-b}%
\begin{align}
M_{nt,D=3}  & =\frac{3g^{4}}{2\pi}\int_{0}^{1}dx\,\frac{e^{-2\sqrt{-L^{2}}%
\sqrt{-s^{2}}\sqrt{(1-x)x}}}{\sqrt{-s^{2}}\sqrt{(1-x)x}}\cos((1-2x)s\cdot
L),\\
M_{nt,D=4}  & =-\frac{8g^{4}}{\pi^{2}}\int_{0}^{1}dx\,K_{0}(\sqrt{-L^{2}}%
\sqrt{-s^{2}(1-x)x}\,)\cos((1-x)s\cdot L)\nonumber\\
& +\frac{2g^{4}}{\pi^{2}}\int_{0}^{1}dx\,K_{0}(2\sqrt{-L^{2}}\sqrt{-s^{2}%
(1-x)x}\,)\cos((1-2x)s\cdot L),
\end{align}
where $K_{0}$ is the modified Bessel function of the second kind. (\ref{np-b})
is complicated and hard to be compared with the right hand side of
(\ref{optical thm}). However, if we focus on the situation for $s^{2}<0$
(spacelike), the right hand side of (\ref{optical thm}) vanishes because
energy-momentum conservation forbids two particles (whether on-shell or
off-shell) with spacelike $s$ to scatter into two real photons.

Assume $s^{2}<0$. For $L^{2}<0$ (spacelike), obviously
(\ref{np-b}) gives $\operatorname{Im}M_{np}=0$ and so unitarity is
satisfied. For $L^{2}>0$ (timelike), however, (\ref{np-b}) leads
to $\operatorname{Im}M_{np}\neq0$ and therefore unitarity is
violated.

\subsection{Scattering with External Photon Legs}

Finally, we study the unitarity for the diagrams with external
photon legs. Consider the vacuum polarization diagram in
Fig.~\ref{fig6}. The amplitude for the nonplanar part is given by
\end{subequations}
\begin{equation}
iM_{np}=-2g^{2}\int\frac{d^{D}l}{(2\pi)^{D}}\frac{\epsilon_{\mu}\epsilon_{\nu
}^{\ast}(k_{\mu}+2l_{\mu})(k_{\nu}+2l_{\nu})}{(l^{2}-m^{2}+i\varepsilon
)((l+k)^{2}-m^{2}+i\varepsilon)}\cos k\cdot L. \label{photon legs}
\end{equation}

The dipole phase $\cos k\cdot L$ in (\ref{photon legs}) is real
and completely decoupled from the internal momentum $l$.
Therefore, the resulting amplitude will not give a new branch cut
and the imaginary part of the amplitude will be given by the same
cutting rule as the ordinary commutative field theory. Thus,
unitarity is always satisfied up to one-loop level no matter $L$
is spacelike, timelike, or even lightlike.

At one-loop level, the same conclusion can be made for $2\gamma
\rightarrow2\gamma$ processes as well. However, the unitarity
might still be violated at the two-loop or higher-loop level since
the dipole phases will be coupled with the internal momenta at the
higher-loop level. The fact that the unitarity is satisfied at
one-loop level whenever the external legs of the diagram are all
photons suggests that some symmetry in the theory saves the
unitarity and should be understood in the limit of the string
theory realization. We defer this issue to our further study.

\section{A Nonlocal Quantum Mechanics}\label{sec:nlqm}
Violation of unitarity also implies that some energy levels get an imaginary
part. This is easiest to analyze for 0+1D systems. In noncommutative
geometry, spacetime must be at least two dimensional, but the nonlocal
dipole theories can be defined with no space and only time. We will now
explore this manifestation of nonunitarity.

We will discuss two theories. The first is a harmonic oscillator with the
nonlocal action
\begin{equation}
S=\frac{1}{2}\int \left\{\dot{x}(t)^{2}+x(t-T)x(t+T)\right\}dt,\qquad x:(-\infty ,\infty
)\mapsto \mathbb{R}  \label{LagNLHO}
\end{equation}%
and the second is a pair of local harmonic oscillators deformed by a
nonlocal interaction term:
\begin{equation}
S=\frac{1}{2}\int \left\{|\dot{z}(t)|^{2}+|z(t)|^{2}+\lambda |z(t)|^{4}-\lambda
|z(t+T)|^{2}|z(t-T)|^{2}\right\}dt,\qquad z:(-\infty ,\infty )\mapsto \mathbb{C}.
\label{LagAHO}
\end{equation}%
Here $t$ is real and $z$ is complex and we set $\hbar =1$, $m=1$ and $\omega
=1$. The second action is the action of a dipole theory with a complex field
$z$ of dipole length $2T$ and an interaction term proportional to $(z\ast
z^{\dagger }-z^{\dagger }\ast z)^{2}.$

How will we find the energy levels? Because of the time nonlocality, a
Hamiltonian formulation is rather cumbersome (see \cite{Woodard:2000bt} for
details). Instead, we will adopt the following approach. We compactify time
on a circle of radius $R$ (in Euclidean time) and expand the partition function
as
\begin{equation}
Z(R)=\sum_{n}C_{n}e^{-2\pi RE_{n}}.  \label{spectrum}
\end{equation}%
Because of nonunitarity, there is no guarantee that $C_{n}=1,$ as we will
see. The energy levels can be read off as the poles of the Fourier transform
of $Z^{\prime }(R)/Z(R)$ as shown below.

\subsection{Nonlocal Harmonic Oscillator}
In the compactified Euclidean time, the action (\ref{LagNLHO}) can be written as%
\begin{equation}
S=\pi R\sum_{n=-\infty}^\infty\left\vert q_{n}\right\vert ^{2}\left\{ \left(\frac{n}{R}\right)^{2}+e^{%
\frac{-2inT}{R}}\right\}
\end{equation}%
in terms of the Fourier transformed modes of $x(t)$: $q_{n}=q_{-n}^{\ast }=%
\frac{1}{2\pi R}\int_{0}^{2\pi R}x(t)e^{-int/R}dt$. The partition function
is then obtained by path integral over the momentum space:%
\begin{equation}
Z(R)\propto \int \prod_{n=-\infty }^{\infty }dq_{n}\ e^{-S}\propto
\int \prod_{n=1}^{\infty }dq_{n}dq_{n}^{\ast }\ e^{-\pi
R\sum_{n=1}^\infty \left\vert q_{n}\right\vert ^{2}\left\{
2\left(\frac{n}{R}\right)^{2}+e^{\frac{-2inT}{R}}+e^{\frac{2inT}{R}}\right\}
}.
\end{equation}%
By $\int_{-\infty }^{\infty }dzdz^{\ast }e^{-\alpha \left\vert
z\right\vert ^{2}}=\pi /\alpha $, the result reads as
\begin{equation}
Z(R)=\frac{1}{2\pi R\prod_{n=1}^{\infty }\left( 1+\frac{R^{2}}{n^{2}}\cos
\frac{2nT}{R}\right) },  \label{partition}
\end{equation}%
where the overall constant factor is fixed such that when $T=0$, $%
Z(R)=1/(e^{\pi R}-e^{-\pi R})=\sum_{n=0}^{\infty }e^{-2\pi R(n+1/2)}$ as it should be.

If $Z(R)$ has an expansion of the form (\ref{spectrum}), then
\begin{equation}
\log Z(R)=\log C_{0}-2\pi E_{0}R+\frac{C_{1}}{C_{0}}e^{-2\pi (E_{1}-E_{0})R}+%
\frac{C_{2}}{C_{0}}e^{-2\pi (E_{2}-E_{0})R}+\cdots ,
\end{equation}%
where $(\cdots )$ are higher exponentials which are products of the $%
e^{-2\pi (E_{n}-E_{0})R}$ terms. The derivative of $\log Z(R)$ gives
\begin{equation}
\frac{Z^{\prime }(R)}{Z(R)}=2\pi \left\{ -E_{0}-\sum_{n=1}^{\infty }\frac{%
C_{n}}{C_{0}}(E_{n}-E_{0})e^{-2\pi (E_{n}-E_{0})R}+\cdots \right\} .
\end{equation}%
The Fourier transform of the equation above is
\begin{equation}
\hat{Z}(\xi )\equiv \int_{0}^{\infty }e^{-2\pi i\xi R}\frac{Z^{\prime }(R)}{%
Z(R)}\,dR=\frac{iE_{0}}{\xi }+\sum_{n=1}^{\infty }\frac{C_{n}}{C_{0}}\frac{%
i(E_{n}-E_{0})}{\xi -i(E_{n}-E_{0})}+\cdots ,\quad \text{with }\operatorname{Im}\xi
<0,  \label{fourier Z}
\end{equation}%
from which the energy spectrum can be read off as the poles of $\hat{Z}(\xi )
$ for $\xi \in \mathbb{C}$.

On the other hand, directly from (\ref{partition}), we get
\begin{equation}
\frac{Z^{\prime }(R)}{Z(R)}=-\frac{1}{R}-\sum_{n=1}^{\infty }\frac{2R\cos
\frac{2nT}{R}+2nT\sin \frac{2nT}{R}}{n^{2}+R^{2}\cos \frac{2nT}{R}}.
\end{equation}%
We can use Poisson summation formula to write
\begin{eqnarray}
& & \frac{Z^{\prime }(R)}{Z(R)} \nonumber \\
&=&-\int_{0}^{\infty }\frac{2R\cos \frac{2xT}{R}%
+2xT\sin \frac{2xT}{R}}{x^{2}+R^{2}\cos \frac{2xT}{R}}\,dx-2\sum_{n=1}^{\infty
}\int_{0}^{\infty }\frac{2R\cos \frac{2xT}{R}+2xT\sin \frac{2xT}{R}}{%
x^{2}+R^{2}\cos \frac{2xT}{R}}\cos (2\pi nx)\,dx  \nonumber \\
&=&-\int_{0}^{\infty }\frac{2\cos 2xT+2xT\sin 2xT}{x^{2}+\cos 2xT}%
\,dx-2\sum_{n=1}^{\infty }\int_{0}^{\infty }\frac{2\cos 2xT+2xT\sin 2xT}{%
x^{2}+\cos 2xT}\cos (2\pi nRx)\,dx.  \label{Z'/Z} \nonumber \\
\end{eqnarray}%
With $\int_{0}^{\infty }\cos (2\pi nRx)e^{-2\pi i\xi R}dR=i\xi /2\pi
(n^{2}x^{2}-\xi ^{2})$, we have the Fourier transform of (\ref{Z'/Z}):%
\begin{eqnarray}
\hat{Z}(\xi ) &=&-\frac{1}{2\pi i\xi }\int_{0}^{\infty }\frac{2\cos
2xT+2xT\sin 2xT}{x^{2}+\cos 2xT}\,dx  \nonumber \\
&&-\sum_{n=1}^{\infty }\int_{-\infty }^{\infty }\frac{2\cos 2zT+2zT\sin 2zT}{%
z^{2}+\cos 2zT}\frac{i\xi }{2\pi (nz-\xi )(nz+\xi )}\,dz.  \label{Zhat}
\end{eqnarray}%
To evaluate the second term of (\ref{Zhat}), consider the contour enclosing
the upper half complex plane. The poles $-\xi /n$ are inside the contour
while the poles $\xi /n$ are outside (remember $\operatorname{Im}\xi <0$). By
residue theorem, the residues at $-\xi /n$ give%
\begin{equation}
\hat{Z}(\xi )=\frac{i}{\xi }E_{0}(T)-\sum_{n=1}^{\infty }\frac{n\cos \frac{%
2\xi T}{n}+\xi T\sin \frac{2\xi T}{n}}{\xi ^{2}+n^{2}\cos \frac{2\xi T}{n}}%
+\cdots ,  \label{fourier Z 2}
\end{equation}%
where%
\begin{equation}
E_{0}(T)\equiv \frac{1}{2\pi }\int_{0}^{\infty }\frac{2\cos 2xT+2xT\sin 2xT}{%
x^{2}+\cos 2xT}\,dx,
\end{equation}%
and $(\cdots )$ are the residues contributed from the singularities which
satisfy $z^{2}+\cos 2zT=0$.

Comparing (\ref{fourier Z 2}) with (\ref{fourier Z}), we find the ground
state energy is $E_{0}=E_{0}(T)$, which reduces to $1/2$, the ground
state energy of local  harmonic oscillators, as $T=0$. Furthermore, when $T=0
$, the pole at $\xi =in$ of (\ref{fourier Z 2}) should match the pole $\xi
=i(E_{n}-E_{0})$ of (\ref{fourier Z}). This gives $E_{n}=E_{0}+n$ as the
spectrum of local harmonic oscillators should be.

When $T\neq 0$, $i(E_{n}-E_{0})$ will be the solutions to%
\begin{equation}
\xi ^{2}+n^{2}\cos \frac{2\xi T}{n},\qquad \operatorname{Im}\xi \geq 0.
\label{dispersion}
\end{equation}%
If we interpret $f=(E_{n}-E_{0})/n=-i\xi /n$ as the frequency of the
classical solution, (\ref{dispersion}) is exactly the same as Equation (56)
in \cite{Woodard:2000bt}, which is derived from the classical equation of
motion of a nonlocal harmonic oscillator. Following \cite{Woodard:2000bt},
the graphical analysis shows $f$ has complex solutions to (\ref{dispersion}%
).\footnote{The complex solutions to the frequency lead to the
instability for classical systems with nonlocality of finite
extent. This is discussed in \cite{Woodard:2000bt}.} Therefore,
some excited energy levels are complex-valued when $T\neq 0$,
which implies the violation of unitarity.

\subsection{Local Harmonic Oscillator with a Nonlocal Perturbation}

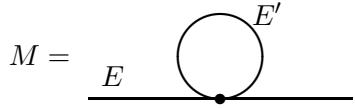
\begin{figure}
\begin{picture}(420,50)

\put(150,20){$M = $}
\put(180,7){\begin{picture}(10,10)
  \put(62,28){$E'$}
  \put(5,5){$E$}
  \thicklines
  \put(0,0){\line(1,0){100}} %
  \put(50,16){\circle{30}} %
  \put(50,0){\circle*{4}} %
  \end{picture}}
\end{picture}
\caption{Self-Energy correction.} \label{fig:SelfEnergy}
\end{figure}

Now we work with the action (\ref{LagAHO}). We treat the action as
a 0+1D field theory and calculate the self-energy correction to
the propagator of $z$. As shown in Fig.~\ref{fig:SelfEnergy}, the
one-loop diagram is given by
\begin{equation}
iM(E)=\frac{\lambda}{2}\int_{0}^{\infty}\frac{dE^{\prime}}{E^{\prime
2}-1+i\varepsilon}\left(
2-e^{i(E-E^{\prime})T}-e^{-i(E-E^{\prime})T}\right)
\end{equation}
with the dipole phases. Under Wick rotation, we get
\begin{align}
M(E_{E})  &
=-\lambda\int_{0}^{\infty}\frac{dE_{E}^{\prime}}{E_{E}^{\prime
2}+1}\left[  \left(  1-\cos((E_{E}-E_{E}^{\prime}\right)  T_{E})\right] \nonumber \\
& =-\frac{\lambda\pi}{2}+\frac{\lambda}{2}\sin(E_{E}T_{E})\left\{
-2\operatorname{Ci}(iT_{E})\sinh
T_{E}+[2\operatorname{Si}(iT_{E})+i\pi]\cosh T_{E}\right\}  ,
\end{align}
where $\operatorname{Ci}(z)$ and $\operatorname{Si}(z)$ are cosine
and sine integrals.
Back to the Minkowski space, we have
\begin{equation}
M(E)=-\frac{\lambda\pi}{2}+i\frac{\lambda}{2}\sin(ET)\left\{
2\operatorname{Ci}(T)\sin T+[\pi-2\operatorname{Si}(T)]\cos
T\right\}
\end{equation}
and $\operatorname{Im}M\neq 0$.

However, the diagram in Fig.~\ref{fig:SelfEnergy} cannot be cut
into two separate pieces as described in Subsection
\ref{subsec:optical theorem}. This means the right hand side of
(\ref{optical thm}) is zero. The cutting rule is thus violated
with $\operatorname{Im}M\neq 0$ on the left hand side of
(\ref{optical thm}), and therefore the theory is nonunitary.

Another way to understand the nonunitarity is to consider the
propagator. The propagator with one-loop correction is given by:
\begin{equation}
G(E) = \frac{1}{E^2 -1 -M(E) +i\varepsilon },
\end{equation}
which has a pole with an imaginary part. In field theory, the
imaginary part of the pole indicates that the particle is unstable
and can decay into lighter particles (see \cite{Peskin:Schroeder}
again). However, $z(t)$ is the only field we have, so the particle
simply decays into nothingness, which implies nonunitarity.

\section{Discussion\label{sec:discussion}}

In this paper, we have found that both $\phi\rightarrow\phi$ and
$2\phi\rightarrow 2\phi$ processes are nonunitary at one-loop
level when the dipole vector $L$ is timelike. On the other hand,
the processes in which the external legs are all photons are
unitary at one-loop level regardless of the signature of $L$.

DFTs can be constructed as an appropriate effective description of
a low energy limit of string theory. In this limit, all the
massive open string states are decoupled from the closed strings
and the relevant degrees of the freedom are the massless open
strings. String theory in this limit can be appropriately
described in terms of DFT and thus the field theory should be
unitary.

However, when the dipole vector is timelike, there is no regime in
which DFT is an appropriate description of string theory and
massive open string states cannot be neglected. This suggests the
violation of unitarity in the field theory when $L$ is timelike as
we have found.

As unitarity of NCSYM can be restored\cite{Seiberg:2000ms,
Gopakumar:2000na, Barbon:2000sg}, by adding an infinite tower of
massive fields, we should have the NCOS completion for DFT to a
unitary theory \cite{Upcoming}.

We also found that some energy levels or the poles of the
propagator of 0+1D quantum mechanical systems with nonlocal
interaction with finite extent in time have an imaginary part and
therefore unitarity is also violated.

\begin{acknowledgments}
It is a pleasure to thank K. Dasgupta, J. Gomis, A. Pasqua, and
M. M. Sheikh-Jabbari for helpful discussions.
This work was supported in part by the Director, Office of
Science, Office of High Energy and Nuclear Physics, of the U.S.
Department of Energy under Contract DE-AC03-76SF00098, and in part
by the NSF under grant PHY-0098840.

\end{acknowledgments}

\end{document}